\documentclass[aps,prl,reprint,showpacs,floatfix,superscriptaddress]{revtex4-1}
\usepackage{graphicx}
\usepackage{amsmath,amsfonts,amscd, physics}
\usepackage{inputenc}
\usepackage{mathrsfs}
\usepackage{mathtools}
\usepackage{stmaryrd}
\usepackage{array}
\usepackage{bm}
\usepackage{setspace}
\usepackage[T1]{fontenc}
\usepackage[american]{babel}
\usepackage{lineno}
\usepackage{csquotes}

\begin{document}
\title{Quantum Fisher Information Bounds on Precision Limits of Circular Dichroism}
\author{J. Wang}
\affiliation{Department of Physics and Astronomy, Texas A\&M University, College Station, Texas 77843, USA}
\author{G. S. Agarwal}
\affiliation{Department of Physics and Astronomy, Texas A\&M University, College Station, Texas 77843, USA}
\affiliation{Institute for Quantum Science and Engineering and Department of Biological and Agricultural Engineering, Texas A\&M University, College Station, Texas 77843, USA}

\begin{abstract}
Circular dichroism (CD) is a widely used technique for investigating optically chiral
molecules, especially for biomolecules. It is thus of great importance that these parameters be
estimated precisely so that the molecules with desired functionalities can be designed. In
order to surpass the limits of classical measurements, we need to probe the system with quantum
light. We develop quantum Fisher information matrix (QFIM) for precision estimates of the
circular dichroism and the optical rotary dispersion for a variety of input quantum states of
light that are easily accessible in laboratory. The Cramer-Rao bounds, for all four chirality parameters are obtained, from QFIM for (a)
single photon input states with a specific linear polarization and for (b) NOON states having
two photons with both either left polarized or right polarized. The QFIM bounds, using quantum
light, are compared with bounds obtained for classical light beams i.e., beams in coherent
states. Quite generally, both the single photon state and the NOON state exhibit superior
precision in the estimation of absorption and phase shift in relation to a coherent source of
comparable intensity, especially in the weak absorption regime. In particular, the NOON state
naturally offers the best precision among the three. We compare QFIM bounds with the error
sensitivity bounds, as the latter are relatively easier to measure whereas the QFIM bounds
require full state tomography. 
\end{abstract}

\maketitle

\section{I. INTRODUCTION}

The estimation of physical quantities is a central theme in scientific experiments and industrial
enterprises. To enable the development of modern metrological appliances and state-of-the-art technology,
devising schemes for improving and optimizing the precision is of critical importance. The core objective of
precision measurements has given rise to the field of quantum estimation which makes use of sophisticated
quantum light sources such as squeezed states \cite{bondurant, teich1989}, entangled photon pairs
\cite{mandel, gisin}, single-photon sources \cite{marocco, michel, beveratos}, and so on. There are,
however, inherent theoretical challenges to extracting full information about any parameter of interest. To
quantify theoretical constraints to parameter estimation,the Fisher information method \cite{cramer, fisher}
is used to obtain a lower bound to the precision of a classical measurement, known as the Cram\'er-Rao
bound. This classical method has been generalized to the quantum formalism \cite{helstrom, holevo, caves2,
davi1, davi2, davi3, braun2, jordan}. The quantum Fisher information (QFI) involving a set of parameters
yields the absolute lower bound to the measurement uncertainties with respect to a specific input state,
which is independent of the measurement setup. With the bulk of quantum resources available, quantum
estimation has been applied to many experiments. In particular, the advent of single-photon
detectors\cite{becker, bachor} has provided the logistic framework for implementing measurements of the QFI.

In this paper, we demonstrate how suitable choices of quantum input, integrated with single-photon detectors
aimed at measuring the QFI, can yield an enhanced estimation of the physical parameters relevant to circular
dichroism (CD). The CD is a well-established technique that studies the differences in light-matter
interaction in an optical medium between the left- and right-circularly polarized components. As a practical
technique, this finds tremendous importance in the study of bio-molecules and other scientific fields. It
has applications in probing tiny molecules, including biological macromolecules, such as proteins, nucleic
acids, carbohydrates, etc. The CD can be used to unveil the secondary structure of a protein, which, in
turn, would shed light on the protein's function \cite{greenfield, whitmore, provencher, sreerama}. More
intricate structural details of biomolecules, like antibodies \cite{cathou68, cathou70, joshi}, can be
investigated through CD than by analyzing the optical rotatory dispersion spectrum. CD of a single cell can
be measured as a function of the position in the cell cycle \cite{bienkiewicz}, and is sensitive to molten globule intermediates which might be involved in the folding process \cite{jennings}. Thus, it can
assess the structure and stability of the protein fragments. Inorganic chiral nanoparticles or quantum dots,
which are expected to work as artificial proteins for chiral catalysis or inhibition of specific enzymes,
have been shown to demonstrate size-dependent CD absorption features \cite{tang}. Interestingly, the
technique can even be observed remotely at astronomical distances, which might prove contributory to the
search of extraterrestrial life \cite{bailey}.

\begin{figure}[h!]
\centering\includegraphics[width=8cm]{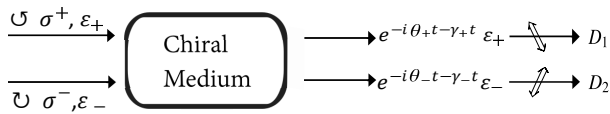}
\caption{Left and right handed circularly polarized light fields undergo differential transmissions through a chiral medium, as characterized by distinct absorption and phase shift rates. The detectors can include polarizers before them.}
\label{Fig1}
\end{figure}

Classical ellipsometry utilizes the polarization of light to study the reflection amplitude and the phase
shift between the reflected and the incident light from a material medium \cite{azzam}. Here, we
contextualize the theory of quantum estimation to the study of CD by measuring the transmission
characteristics of a chiral medium. As shown in Fig.\ref{Fig1}, four parameters characterize this chiral
interaction process: the dimensionless net absorption coefficients for the two circularly polarized light waves $(\alpha _{+}, \alpha
_{-})$, and the corresponding dimensionless net phase shifts $(\phi _{+}, \phi _{-})$. Equivalently, one can treat the sums
and differences of these pairs, by introducing a new parameter family, $X_{s}=(\alpha _{+}+\alpha _{-})/2$,
$X_{d}=(\alpha _{+}-\alpha _{-})/2$, $\varSigma=\phi _{+}+\phi _{-}$ and $\triangle =\phi _{+}-\phi
_{-}$, where we assume that $\alpha _{+}>\alpha _{-}$ and $\phi _{+}>\phi _{-}$. All these parameters are functions of the longitudinal dimension of the medium. 
The standard ellipsometry provides measurements of the four chirality parameters: $\alpha _{+}, \alpha _{-},
\phi _{+}, \phi _{-}$. In particular circular dichroism is given
by the parameter $X_{d}$ and the optical
rotary dispersion (ORD) is given by $\triangle$. The Stokes polarimetry can also be used to obtain complete
polarization state of the output beam. The classical results on parameter estimation are limited by the
standard quantum noise limit which can be surpassed by the use of quantum light such as squeezed
light\cite{caves}. Clearly we need to use quantum light for precision estimates of the chirality parameters
for both CD and ORD. For quantum inputs we need to do complete state tomography\cite{kwait, agarwal10}. 
A scheme for ellipsometry with twin photons produced by the spontaneous parametric down conversion (SPDC)
was introduced as a self-referenced method without any calibrated source or a detector \cite{teich1, teich2,
toussaint}, which manifested a sizable improvement through the use of entangled quantum states. The method
was based on the intensity correlations between the output twin beams. But the sensitivity of the
measurement scheme was not discussed. Several other studies have shown advantages of using squeezed light
with tailored beams; SPDC photons in ellipsometry \cite{huber, kolkiran}. The ellipsometry with classically
correlated beams was discussed in \cite{setala}.

We briefly outline the organization of the paper here. In Sec. II, we summarize the key features of the
QFIM. In Sec. III, we introduce the master equation to obtain the quantum state of the output field in terms
of the input state. The master equation is needed as a chiral system is an open system. In sections IV, V,
and VI, we apply the QFIM method and obtain the Cram\'er-Rao bounds for the uncertainties and the
correlations of chiral parameters with coherent light, a linearly polarized single-photon Fock state, and a
NOON state produced in a collinear type-II SPDC process, respectively. We compare the obtained bounds for
the different states and plot them against the absorption sensitivities of a standard intensity measurement,
thereby illustrating the remarkably precision offered by the NOON state in the estimation of circular
dichroism. In section VII, we highlight how the sensitivity in the determination of the relative phase shift
is doubled on using the NOON state as compared to a single-photon Fock state.

\section{II. SUMMARY OF KEY FEATURES OF THE QUANTUM FISHER INFORMATION MATRIX}

In the estimation of an unknown parameter $X$, the measurement uncertainty or sensitivity of $X$ is always
bounded as $\delta X\geq \sqrt{F_{Q}^{-1}(X)}$, which constitutes the Cram\'er-Rao bound for a
single-parameter measurement. Here, the QFI $F_{Q}(X)$ is defined by $F_{Q}(X)=Tr[\rho L^{2}]$, where the
symmetric logarithmic derivative (SLD) matrix $L$ follows from the equation $\frac{\partial \rho }{\partial
X}=\frac{1}{2}(L\rho +\rho L)$, with $\rho$ being the density matrix of the system. In practice, one needs
to determine several parameters which characterize the physical system, each of which suffers from similar
kinds of sensitivity constraints. If we consider a comprehensive set of parameters $\{X_{i}\}$, for
$i=1,2,...,N$, the SLD for each parameter would be obtained from the equation $\frac{\partial\rho }{\partial
X_{i}}=\frac{1}{2}(L_{i}\rho +\rho L_{i})$. The generalized quantum Fisher information, now expressed as a
matrix QFIM, is constructed from elements given by 
\begin{align}
F_{i,j}(\{X_{i}\})=\frac{1}{2}Tr[\rho(L_{i}L_{j}+L_{j}L_{i})].
\end{align}
The whole SLD matrix would generally contribute to the sensitivity of determining a specific parameter. The
modified Cram\'er-Rao bound $Cov(X_{i},X_{j})\geq {[F^{-1}(\{X_{i}\})]_{ij}}$ then yields the lower bound on
the sensitivity of $X_{i}$ to be 
\begin{align}
\delta X_{j}\geq \sqrt{\lbrack F^{-1}(\{X_{i}\})]_{jj}}.
\end{align}
Note that the sensitivity bound is determined in terms of the diagonal element of the inverse matrix
$F^{-1}$, which is contingent on all the elements of the QFIM. For simplicity, we can introduce the
sensitivity bound matrix (SQFIM) defined by $S_{i,j}=\sqrt{(F^{-1})_{i,j}}$. Based on this formulation, we
could estimate the sensitivities of chiral parameters in a standard ellipsometric setup for various input
states of light. Following the QFI formalism, the sensitivity bounds for absorption rate of a single mode of
the field have been evaluated with both a Gaussian state \cite{paris, braun1} and non-Gaussian states such as a Fock (photon number)
state \cite{souza, jiaxuan}. The Fock state has turned out to be an optimal choice for the estimation. At zero
temperature, the sensitivity derived from ordinary intensity fluctuations saturates the QFI bound for the
single absorption mode with a Fock state input. For an output intensity of the form $N(\alpha )$, the error
sensitivity of the parameter $\alpha$ is determined in a single experiment via the relation
$\widetilde{\delta \alpha}=\frac{\delta N}{\partial N/\partial \alpha}$, where $(\delta N)^{2}= \langle
N^{2}\rangle-\langle N \rangle^{2}$. However, when multiple unknown parameters are connected to the
propagation dynamics of light, the estimation uncertainty of any particular parameter is not independent of
the others, and is related, as would be seen, by the QFI matrix (QFIM) \cite{chen, liu, datta, ioannou}. A pertinent question is: are there measurements which can optimize the bounds for two parameters at the same time. Crowley et al\cite{crowley} investigated the
measurements of both phase and absorption of a single mode, and found that there is a trade off between the simultaneous measurement of phase and absorption, i.e. if scheme is optimised for absorption measurements, then it is not optimal for phase.
In this paper, we focus on initial states of fields which are now easily accessible in the laboratory, rather than studying the question when a measurement is optimal.

\section{III. THE EVOLUTION OF DENSITY MATRIX OF THE FIELDS}

The transport properties of electromagnetic field in a medium follow from Maxwell's equations and are
determined by the refractive index of the medium. At the classical level, the absorption and the phase shift
of light through the medium are encoded in the output field amplitude which is related to the input
amplitude as $\varepsilon_{out}=e^{-i\theta t-\gamma t}\varepsilon_{in}$. In the quantum mechanical prescription, the evolution
of the system is described by the master equation for the density matrix $\rho$ of light. For a single-mode
input, the phase shift by itself is described as a unitary process via the von-Neumann equation
$\frac{\partial\rho}{\partial t}=-i[H,\rho]$, where $H=\theta a^{\dagger}a$, and $a$ denotes the Bosonic
annihilation operator for the light field. The process of absorption needs recourse to the master equation
for a damped field, i.e.,
\begin{equation}
\frac{\partial\rho}{\partial t}=-i\theta[a^{\dagger }a,\rho]-\gamma(\rho a^{\dagger}a-2a\rho a^{\dagger}+a^{\dagger}a\rho)\,,  \label{3mast}
\end{equation}
where the time dependence also can be considered as medium-length dependence as $l=tc$.
Owing to the independence of the two processes, the two dynamical equations can be superposed to yield a
simple solution akin to the classical result, $\expval{a(t)}=e^{-i\theta t-\gamma t}\expval{a(0)}$. More
generally, for a chiral medium, in which the photon transfer is sensitive to the input polarization, the
full master equation would be expressed as
\begin{align}
\frac{\partial \rho }{\partial t}=& -i\theta _{+}[a_{+}^{\dagger }a_{+},\rho
] -\gamma _{+}(\rho a_{+}^{\dagger }a_{+}-2a_{+}\rho a_{+}^{\dagger
}+a_{+}^{\dagger }a_{+}\rho )  \notag \label{master}  \\
& -i\theta _{-}[a_{-}^{\dagger }a_{-},\rho ] -\gamma _{-}(\rho a_{-}^{\dagger }a_{-}-2a_{-}\rho a_{-}^{\dagger}+a_{-}^{\dagger }a_{-}\rho )\,.
\end{align}

\noindent Here, $\gamma _{\pm }=-ln(1-\alpha _{\pm })/(2t)$ and $\theta _{\pm }=\phi _{\pm }/t$ are the
damping and phase-shift rates pertaining to the two circular polarization directions, $a_{\pm }$ and $a_{\pm
}^{\dagger }$ are the annihilation and creation operators of photons obeying the commutation relation
$[a_{\pm
},a_{\pm }^{\dagger }]=1$, with $a_{\pm}=\frac{1}{\sqrt{2}}(a_{H}\pm ia_{V})$. The four parameters $\alpha
_{\pm }, \phi _{\pm }$ were introduced earlier in Sec. I. For a known initial state $\rho(0)$, the density
matrix $\rho (t)$ at time $t$ can be straightforwardly calculated from the master equation. Subsequently,
the necessary SLDs: $L_{d}$, $L_{s}$, $L_{\triangle }$, and $L_{\varSigma}$ can be computed at any arbitrary
time in terms of the derivatives $\frac{\partial \rho (t)}{\partial X_{d}}$, $\frac{%
\partial \rho (t)}{\partial X_{s}}$, $\frac{\partial \rho (t)}{\partial X_{\triangle }}$, $\frac{\partial
\rho (t)}{\partial X_{\varSigma}}$. In the next few sections, we would present explicit solutions for
several input states of interest, and demonstrate how quantum sources outperform classical light by
furnishing improved sensitivity bounds. Specifically, we would consider a coherent state, a single-photon
state, and a two-photon entangled state to establish this result.

\section{IV. BOUNDS ON THE MEASUREMENTS OF CD PARAMETERS WITH CLASSICAL LIGHT \bm{$|\alpha_{H},\beta_{V}\rangle$}}

The simplest method of precision measurement of CD parameters is via the estimation of intensity
fluctuations. Classically, the uncertainty of an ellipsometric parameter $X$ translates into a fluctuation
in the output intensity which is connected to the former through the propagation of uncertainty. For a
functional dependence $I_{out}=I_{out}(X, I_{in})$, we have $\delta X=\frac{\delta I_{out}}{\partial I_{out}
/ \partial X}$. In the case of classical light, we prove that the magnitudes of $\delta X_{d}$
and $\delta X_{s}$ obtained from intensity measurements saturate the Cram\'er-Rao bound. Classical light
with two polarizations is described by a coherent state $\ket{\alpha}_{H}\ket{\beta}_{V}$, which can be
recast as

\begin{equation}
|\psi (0)>=\ket{\frac{\alpha +i\beta }{\sqrt{2}}}_{+}\ket{\frac{\alpha -i\beta }{\sqrt{2}}}_{-}\,,  \label{4phi0}
\end{equation}

\noindent in the basis of eigenstates of $a_{\pm }$. Since this is a product state, the solution to the master equation in this case reads $\rho (t)=\ket{\psi (t)}\bra{\psi (t)}$, where 

\begin{align}
|\psi (t)>&=\ket{(\alpha+i\beta)\sqrt{\frac{1-\alpha_{+}}{2}}e^{-i\theta _{+}t}}_{+} \notag \\
&\otimes \ket{(\alpha-i\beta)\sqrt{\frac{1-\alpha_{-}}{2}}e^{-i\theta _{-}t}}_{-}, \phi_{\pm}=\theta _{\pm}t\,.
\label{4phit} 
\end{align}

\noindent Thus, the coherent input goes over into a coherent output, albeit with modified amplitude and
phase. Using this solution, we first calculate sensitivities as obtained by intensity fluctuations. As
sketched in Fig. \ref{Fig1}, the input light first goes through the measured sample, and then, a
polarization analyzer, so that we can detect a certain polarized output. For the pair of measured
intensities $I_{\pm}$ corresponding to the two polarizations, we have the absorption difference
$X_{d}=(I_{+}-I_{-})/N_{0}$, and the net absorption $X_{s}=1-(I_{+}+I_{-})/N_{0}$, where
$N_{0}$ is the input photon number, as for simplicity, we set the relative phase between $\alpha$ and $\beta$ zero , thus $|\alpha \pm i \beta|^{2}=|\alpha|^{2}+|\beta|^{2}$. The sensitivities then unfold as

\begin{equation}
\widetilde{\delta X_{d}}=\sqrt{(\delta I_{+})^{2}+(\delta I_{-})^{2}-2Cov(I_{+}, I_{-})}/N_{0}  \,,
\end{equation} 
\begin{equation}
\widetilde{\delta X_{s}}=\sqrt{(\delta I_{+})^{2}+(\delta I_{-})^{2}+2Cov(I_{+}, I_{-})}/N_{0}  \,,
\end{equation} 

\noindent both of which reduce to 
\begin{align}
\widetilde{\delta X_{d}}=\widetilde{\delta X_{s}}=\sqrt{\frac{1-X_{s}}{N_{0}}}.
\end{align} 

Next, we calculate the bound by following the approach outlined in Sec II. We obtain the corresponding SLDs at time $t$ as 
\begin{equation}
L_{d}=\frac{1}{(1-\alpha_{+})}a_{+}^{+}a_{+}-\frac{1}{(1-\alpha_{-})}a_{-}^{+}a_{-}\,,  \label{1ld}
\end{equation}
\begin{equation}
L_{s}=-\frac{1}{(1-\alpha_{+})}a_{+}^{+}a_{+}-\frac{1}{(1-\alpha_{-})}a_{-}^{+}a_{-}+N_{0}\,,  \label{1ls}
\end{equation}
\begin{equation}
L_{\varSigma}=-2i[G_{\varSigma},\rho ]\,,\,\,G_{\varSigma}=%
(a_{+}^{+}a_{+}+a_{-}^{+}a_{-})\,,  \label{1lsigma}
\end{equation}%
\begin{equation}
L_{\triangle }=-2i[G_{\triangle },\rho ]\,,\,\,G_{\triangle }=%
(a_{+}^{+}a_{+}-a_{-}^{+}a_{-})\,.  \label{1ldelta}
\end{equation}

\noindent The derivation of these SLDs is shown in Appendix A. This leads us to the QFIM 
\begin{equation}
F=\left[ 
\begin{array}{cccc}
F_{dd} & F_{ds} &  &  \\ 
F_{ds} & F_{ss} &  &  \\ 
&  & F_{\triangle \triangle } & F_{\varSigma\triangle } \\ 
&  & F_{\varSigma\triangle } & F_{\varSigma\varSigma}%
\end{array}%
\right] \,,  \label{1F}
\end{equation}

\noindent where 
\begin{align*}
&F_{dd}=F_{ss}=\frac{N_{0}(1-X_{s})}{%
(1-X_{s})^{2}-X_{d}^{2}}\,, \notag \\
&F_{\triangle\triangle}=F_{\varSigma\varSigma}=N_{0}(1-X_{s})\,,  \notag \\
&F_{ds}=-\frac{N_{0} X_{d}}{%
(1-X_{s})^{2}-X_{d}^{2}}\,,  
F_{\varSigma\triangle}=-N_{0} X_{d}\,,  \notag \\
\end{align*}

\noindent The empty spaces within the matrix in (14) indicate null matrices, the derivation of which is shown in Appendix B. Being a block-diagonal matrix,
its inverse $F^{-1}$ also possesses a block-diagonal structure, implying that on using coherent state as an
input, the Cram\'er-Rao bounds for the absorption and the phase shift would be independent of each other.
The obtained sensitivity bounds are listed below:

\begin{equation}
\delta X_{d,min}=\delta X_{s,min}=\sqrt{\frac{1-X_{s}}{N_{0}}}\,,  \label{1a}
\end{equation}
\begin{equation}
Cov(X_{d},X_{s})=\frac{X_{d}}{N_{0}}\,,\label{1acov}
\end{equation}
\begin{equation}
\delta \varSigma_{min}=\delta \triangle_{min}=\sqrt{\frac{1}{%
N_{0}}\frac{1-X_{s}}{(1-X_{s})^{2}-X_{d}^{2}}}\,,\label{1phase}
\end{equation}
\begin{equation}
Cov(\varSigma,\triangle)=\frac{1}{N_{0}}\frac{X_{d}%
}{(1-X_{s})^{2}-X_{d}^{2}}\,.\label{1phasecov}
\end{equation}

It follows immediately that the sensitivities obtained via intensity measurements coincide with the Cram\'er-Rao bounds \eqref{1a} entailed by the QFIM method, and thus sensitivity bounds saturate Cram\'er-Rao bounds.

\begin{figure*}[tb]
\centering\includegraphics[width=13cm]{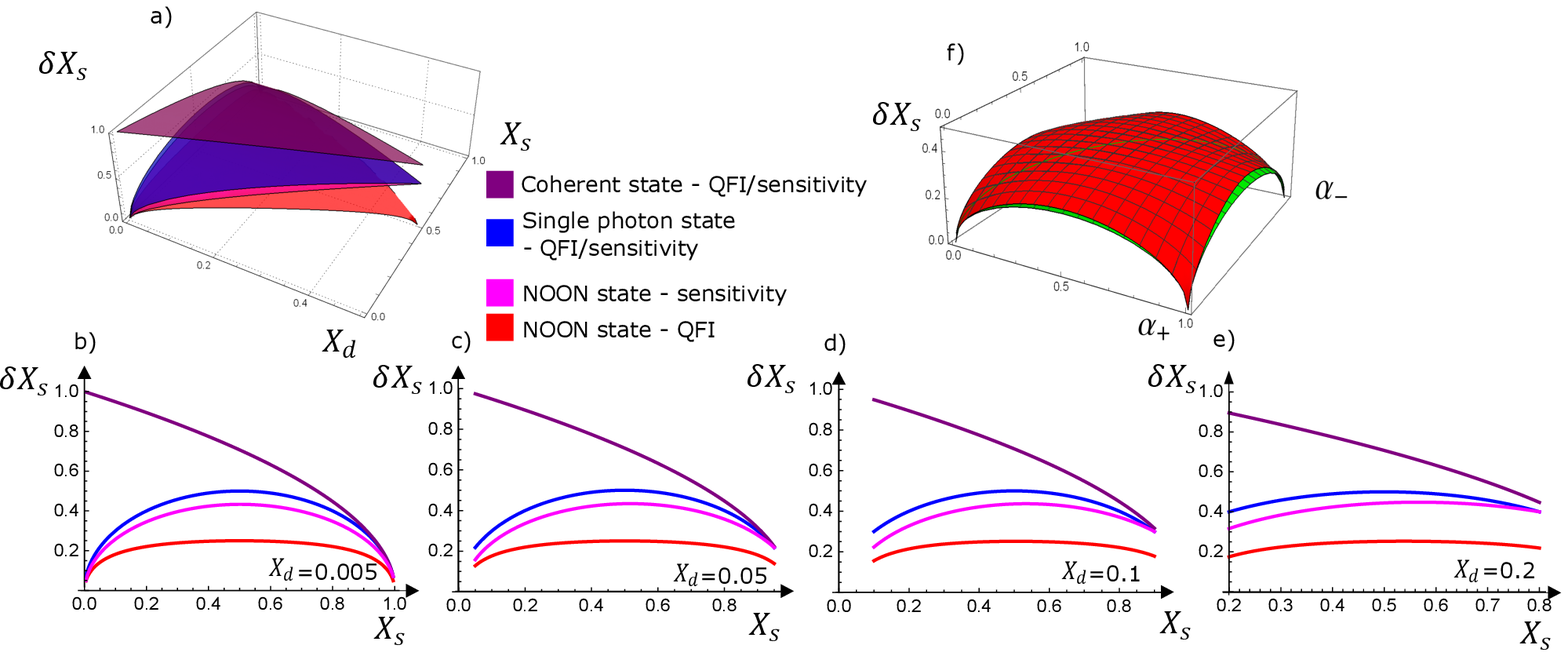}
\caption{a). $\delta X_{s}$ as a function of $X_{d}$ and $X_{s}$. b-e).  For a clearer vision, we also plot $\delta X_{s}$ as a function of $X_{s}$ for different values of $X_{d}$ : b) $X_d=0.005$, c) $X_d=0.05$, d) $X_d=0.1$, and e) $X_d=0.2$. In each plot, the purple line is $\delta X_{d}$ obtained for a coherent state $\ket{\alpha_{H},0_{V}}$, the blue line
is obtained from the state $\ket{1_{H},0_{V}}$, the pink line is
obtained from the intensity measurement with $\ket{1_{H},1_{V}}$, while the red line is obtained from the QFIM of $\ket{1_{H},1_{V}}$. f). $\protect\delta X_{s}$ as a function of $\alpha_{+}$ and $\alpha_{-}$, in which the red (green) curve is obtained from the QFIM of $\ket{1_{H},1_{V}}$ ($\ket{1_{+},1_{-}}$).}
\label{Fig2}
\end{figure*}

\section{V. BOUNDS ON THE MEASUREMENTS OF CD PARAMETERS WITH QUANTUM LIGHT: SINGLE-PHOTON STATE \bm{$|1_{H},0_{V}\rangle$}}

Here, we demonstrate that a single-photon Fock state provides better sensitivity than the coherent state in
the measurement of the absorption rate with a minimum uncertainty in the input photon number, thereby
yielding a definite advantage in estimating the chiral coefficients. Taking the direction of the single
photon's polarization as
horizontal, the input state reads $\ket{1}_{H}\ket{0}_{V}=\frac{1}{\sqrt{2}}%
(\ket{1}_{+}\ket{0}_{-}+\ket{0}_{+}\ket{1}_{-})$. 
The sensitivity obtained from the intensity measurement can be similarly derived via the method invoked in
section IV. With the input field $\ket{1}_{H}\ket{0}_{V}$, the corresponding expressions stand as
\begin{equation}
\widetilde{\delta X_{d}}=\sqrt{1-X_s-X_{d}^{2}}\,, \label{2d0}
\end{equation}
\begin{equation}
\widetilde{\delta
X_{s}}=\sqrt{(1-X_{s})X_{s}}\,.  \label{2s0}
\end{equation}
\noindent Next, we study the bound obtained from the QFIM.

From the master equation, we obtain the
density matrix at time $t$ as
\begin{align}
\rho (t)& =\frac{1}{2}(1-\alpha_{+})\ket{1_{+},0_{-}}\bra{1_{+},0_{-}} \notag \\
&+\frac{1}{2}(1-\alpha_{-})\ket{0_{+},1_{-}}\bra{0_{+},1_{-}}  \notag \label{2density} \\
& +\frac{e^{i\triangle }}{2}\sqrt{(1-\alpha_{+})(1-\alpha_{-})%
}\ket{1_{+},0_{-}}\bra{0_{+},1_{-}} \notag \\
&+\frac{e^{-i\triangle }}{2}\sqrt{(1-\alpha_{+})(1-\alpha_{-})%
}\ket{0_{+},1_{-}}\bra{1_{+},0_{-}}  \notag \\
&+X_{s}\ket{0_{+},0_{-}}\bra{0_{+},0_{-}}\,. 
\end{align}

\noindent where the time-dependence (length-dependence) is implicit in the variables $X_s$, $X_d$ and
$\Delta$. Fock states have no absolute phase; thus, the sum of two phase
shifts $\varSigma$ doesn't appear in the equation. Nevertheless, we can study the absorption and the
phase-shift difference. The $L_{d}$, $L_{s}$, and $L_{\triangle}$ are $3\times 3$ matrices.  A calculation
of the SLDs from
the density matrix yields 

\begin{equation}
L_{d}=diag(\frac{-1}{1-\alpha_{+}},\frac{1}{1-\alpha_{-}},0)\,,
\end{equation}%
\begin{equation}
L_{s}=diag(\frac{-1}{1-\alpha_{+}},\frac{-1}{1-\alpha_{-}},\frac{1}{X_{s}})\,,
\end{equation}%
\begin{equation}
L_{\triangle }=\frac{2\sqrt{(1-\alpha_{+})(1-\alpha_{-})}}{(1-\alpha_{+})+(1-\alpha_{-})}\left[
\begin{array}{ccc}
& ie^{i\triangle } &\\ 
-ie^{-i\triangle } & &\\
& & 0
\end{array}%
\right] \,.
\end{equation}

\noindent The notation $diag(a_{1}, ... , a_{N})$ is the $N \times N$ diagonal matrix whose entries are the
$N$ elements $a_{1}, ... , a_{N}$. The SLDs for the absorption rates are diagonal. We then find the SQFIM to be 
\begin{equation}
F=\left[ 
\begin{array}{ccc}
F_{dd} & F_{ds} &  \\ 
F_{ds} & F_{ss} &  \\ 
 &  & F_{\triangle \triangle }%
\end{array}%
\right] \,,  \label{2F}
\end{equation}

\noindent where 
\begin{align*}
&F_{dd}=\frac{1-X_{s}}{(1-\alpha_{+})(1-\alpha_{-})}%
\,,  \notag \\
&F_{ss}=\frac{1-X_{s}}{(1-\alpha_{+})(1-\alpha_{-})}+%
\frac{1}{X_{s}}\,,  \notag \\
&F_{\triangle\triangle}=\frac{2(1-\alpha_{+})(1-%
\alpha_{-})}{(1-\alpha_{+})+(1-%
\alpha_{-})}\,,  \notag \\
&F_{ds}=\frac{X_{d}}{(1-\alpha_{+})(1-\alpha_{-})}\,.
\notag \\
\end{align*}

\noindent This is also block-diagonal as the phase shift is not related with absorption in this case. This leads to the following bounds:

\begin{equation}
\delta X_{d,min}=\sqrt{1-X_s-X_{d}^{2}}\,, \label{2d}
\end{equation}
\begin{equation}
\delta X_{s,min}=\sqrt{(1-X_{s})X_{s}}\,,  \label{2s}
\end{equation}
\begin{equation}
Cov(X_{d},X_{s})=-X_{s}X_{d}\,,\label{1sdcorr}
\end{equation}
\begin{equation}
\delta \triangle_{min}=\sqrt{\frac{(1-\alpha_{+})+(1-%
\alpha_{-})}{2(1-\alpha_{+})(1-%
\alpha_{-})}}\,,\label{2phase}
\end{equation}

\noindent  A plot of $\delta X_{s}$ in Fig. \ref{Fig2} brings out the advantage of using the single-photon
Fock state as an input compared to the coherent
state. Clearly, the single-photon state renders a notable improvement in comparison to the coherent state in
the weak absorption regime. It stands to the intuition that an input state with lower fluctuation in the
photon number yields a better sensitivity. As $X_{s}\rightarrow 0$, $\delta
X_{s}$ becomes vanishingly small, while the corresponding sensitivity bound for a coherent state levels off
to 1 when a mean photon number of unity is considered.
However, improvements in the estimation sensitivity of the chiral absorption-rate difference $\delta X_{d}$ are
not as substantial as that for either of the absorption rates $\alpha_\pm$. This sensitivity can be improved by the
administration of a two-photon entangled state as an input, as we illuminate in the next section.

Note that the sensitivity obtained from intensity fluctuations reach the lower bounds
$\delta X_{d,min}$ and $\delta X_{s,min}$ expressed in \eqref{2d} and \eqref{2s}. This shows that the simple intensity measurements in this case are already optimal.

\begin{figure*}[tb]
\centering\includegraphics[width=13cm]{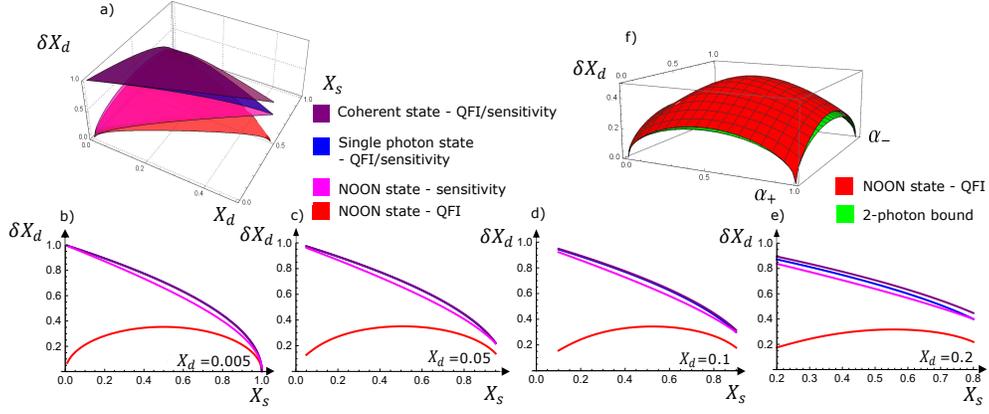}
\caption{a). $\protect\delta X_{d}$ as a function of $X_{d}$ and $X_{s}$. b-e).  For 
a clearer vision, we also plot $\protect\delta X_{d}$ as a function of $X_{s}$ for different
values of $X_{d}$ : b) $X_d=0.005$, c) $X_d=0.05$, d) $X_d=0.1$, and e) $X_d=0.2$. In each plot, the purple line is $\protect\delta X_{d}$
obtained for a coherent state $\ket{\protect\alpha_{H},0_{V}}$, the blue line
is obtained from the state $\ket{1_{H},0_{V}}$, the pink line is
obtained from the intensity measurement with $\ket{1_{H},1_{V}}$, while the red line is obtained from the QFIM of $\ket{1_{H},1_{V}}$. f). $\protect\delta X_{d}$ as a function of $\alpha_{+}$ and $\alpha_{-}$, in which the red (green) curve is obtained from the QFIM of $\ket{1_{H},1_{V}}$ ($\ket{1_{+},1_{-}}$).}
\label{Fig3}
\end{figure*}

\section{VI. BOUNDS ON THE MEASUREMENTS OF CD PARAMETERS WITH QUANTUM LIGHT: NOON STATE \bm{$\frac{1}{\sqrt{2}}(|2_{+},0_{-}\rangle-|0_{+},2_{-}\rangle)$} }

Using a type-II SPDC, one can generate entangled photon pairs with perpendicular polarizations. We choose
the input state to be $\ket{1}_{H}\ket{1}_{V}$. Transformed into the ${\pm}$ basis, it becomes
$\frac{1}{\sqrt{2}}(\ket{2_{+},0_{-}}-\ket{0_{+},2_{-}})$, which embodies a typical two-photon NOON state. 
In this case, the intensity measurement yields a sensitivity of 

\begin{equation}
\widetilde{\delta X_{d}} =\frac{1}{2} \sqrt{(1-\alpha_{+})+(1-\alpha_{-})+2(1-\alpha_{+})(1-\alpha_{-})} \,.
\end{equation}
\begin{equation}
\widetilde{\delta X_{s}} =\frac{1}{2} \sqrt{(1-\alpha_{+})+(1-\alpha_{-})-2(1-\alpha_{+})(1-\alpha_{-})} \,.
\end{equation}

On solving the master equation for this input, we obtain the evolved density matrix as 

\begin{align}
\rho (t)& =\frac{1}{2}(1-\alpha_{+})^{2}\ket{2_{+},0_{-}}\bra{2_{+},0_{-}}\notag \\
 &+\frac{1}{2}(1-\alpha_{-})^{2}\ket{0_{+},2_{-}}\bra{0_{+},2_{-}}  \notag \label{3density} \\
& -\frac{1}{2}e^{i2\triangle }(1-\alpha_{+})(1-\alpha_{-})\ket{2_{+},0_{-}}\bra{0_{+},2_{-}} \notag \\
& -\frac{1}{2}e^{-i2\triangle }(1-\alpha_{+})(1-\alpha_{-})\ket{0_{+},2_{-}}\bra{2_{+},0_{-}} \notag \\
& +\alpha_{+}(1-\alpha_{+})\ket{1_{+},0_{-}}\bra{1_{+},0_{-}} \notag \\
 &+\alpha_{-}(1-\alpha_{-})\ket{0_{+},1_{-}}\bra{0_{+},1_{-}} \notag \\
&+\frac{1}{2}(\alpha_{+}^{2}+\alpha_{-}^{2})\ket{0_{+},0_{-}}\bra{0_{+},0_{-}}\,,  \notag \\
\end{align}

\noindent along with the SLDs 
\begin{align}
L_{d}=diag(\frac{-2}{1-\alpha_{+}},\frac{2}{1-\alpha_{-}},\frac{1-2\alpha_{+}}{\alpha_{+}(1-\alpha_{+})},\notag \\
-\frac{1-2\alpha_{-}}{\alpha_{-}(1-\alpha_{-})}),\frac{2X_{d}}{X_{s}^{2}+X_{d}^{2}})\,,
\end{align}
\begin{align}
L_{s}=diag(\frac{-2}{1-\alpha_{+}},\frac{-2}{1-\alpha_{-}},\frac{1-2\alpha_{+}}{\alpha_{+}(1-\alpha_{+})},\notag \\
\frac{1-2\alpha_{-}}{\alpha_{-}(1-\alpha_{-})},\frac{2X_{s}}{X_{s}^{2}+X_{d}^{2}})\,,
\end{align}
\begin{align}
L_{\triangle }&=\frac{4(1-\alpha_{+})(1-\alpha_{-})}{(1-\alpha_{+})^{2}+(1-\alpha_{-})^{2}} \notag \\
&\times diag(\left( 
\begin{array}{cc}
& ie^{i2\triangle } \\ 
-ie^{-i2\triangle } & 
\end{array}%
\right) , 0,0,0)\,.
\end{align}

\noindent

One can see that $L_{\triangle}$ has a similar form as for the single-photon input state, but the dependence on $\triangle $ is increased by 2. This results in a two-fold enhancement of the sensitivity in $\triangle $. In particular, we find that the QFIM is given by 
\begin{equation}
F=\left[ 
\begin{array}{ccc}
F_{dd} & F_{ds} &  \\ 
F_{ds} & F_{ss} &  \\ 
 &  & F_{\triangle \triangle }%
\end{array}%
\right] \,,  \label{3F}
\end{equation}

\noindent where 
\begin{align*}
&F_{dd}=4+\frac{(1-2\alpha_{+})^{2}}{\alpha_{+}(1-\alpha_{+})}+\frac{(1-2\alpha_{-})^{2}}{\alpha_{-}(1-\alpha_{-})}+\frac{4X_{d}^{2}}{X_{s}^{2}+X_{d}^{2}}\,,  \notag
\\
&F_{ss}=4+\frac{(1-2\alpha_{+})^{2}}{\alpha_{+}(1-\alpha_{+})}+\frac{(1-2\alpha_{-})^{2}}{\alpha_{-}(1-\alpha_{-})}+\frac{4X_{s}^{2}}{X_{s}^{2}+X_{d}^{2}}\,,  \notag
\\
\end{align*}

\begin{align*}
&F_{\triangle\triangle}=\frac{8(1-\alpha_{+})^{2}(1-\alpha_{-})^{2}%
}{(1-\alpha_{+})^{2}+(1-\alpha_{-})^{2}}\,,  \notag \\
&F_{ds}=\frac{4 X_{s}X_{d}}{X_{s}^{2}+X_{d}^{2}}+\frac{(1-2 \alpha_{+})^{2}}{\alpha_{+}(1-\alpha_{+})}-\frac{(1-2 \alpha_{-})^{2}}{\alpha_{-}(1-\alpha_{-})}\,.  \notag
\\
\end{align*}

\noindent

\begin{figure}[tb]
\centering\includegraphics[width=7cm]{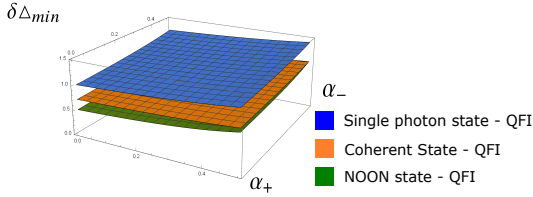}
\caption{$\protect\delta \triangle_{min}$ as a function of $\alpha_{+}$ and $\alpha_{-}$. b-e). The blue curve
is obtained from the state $\ket{1_{H},0_{V}}$, which is the same with a coherent state $\ket{\protect\alpha_{H},\beta_{V}}$ with an average photon number of 1. The orange curve is 
obtained from the QFIM of a coherent state $\ket{\protect\alpha_{H},\beta_{V}}$ with an average photon number of 2. The green curve is obtained from the state $\ket{1_{H},1_{V}}$.}
\label{Fig4}
\end{figure}

The formulae for SQFIM and the resulting sensitivity bounds are too complicated and not very insightful.
Figures \ref{Fig2} and \ref{Fig3} capture the essential features. Clearly,  than intensity-fluctuation
measurements, a direct measurement of the QFI, predicts a much better sensitivity. Further, the QFI method, in
this case, would grant more precise information compared to the single-photon input, especially in the weak
absorption region for $\delta X_{d}$ (Fig. \ref{Fig3}). As $X_{s}\rightarrow 0$, $\delta X_{d}\rightarrow 0$
for the NOON state input, while this uncertainty approaches $1$ for a single-photon input. Using the idea that Fock states are optimal for absorption measurements, we can obtain $\delta X_{d}$ and $\delta X_{s}$ assuming the input state is $\ket{1_{+},1_{-}}$ and get a bound for this particular two photon input state $\sqrt{\alpha_{+}(1 - \alpha_{+}) + \alpha_{-}(1 - \alpha_{-})}/2$ \cite{ioannou}, which is shown as the green curve in Fig. 2. f) and Fig. 3. f). It can be seen from the figures that it is almost the same as our bound obtained from the QFIM of the NOON state. Considering all
of these aspects, we conclude that the use of the SPDC-generated NOON state would be considerably
advantageous in the estimations of both net absorption $X_{s}$ and absorption difference $X_{d}$ in weakly
absorbing samples.

We have shown that the bounds stipulated by the QFIM are lower than the sensitivities obtained by intensity
measurements for the NOON state. The whole density matrix here bears
on the QFIM, and we need to measure all the coefficients in the density matrix. In the equations
\eqref{2density} and \eqref{3density}, the diagonal terms can be measured by single photon detectors. And by
means of projective measurements introduced in section VII, we can infer the off-diagonal terms via
fidelity estimations. The magnitudes and the phases of these off-diagonal terms would be given by the
respective amplitudes and frequencies of the fringes.

There is also advantage of using the NOON state to obtain a better achievable sensitivity of relative phase $\protect\delta \triangle_{min}$. As shown in Figure \ref{Fig4}, at the weak absorption limit $X_s \to 0$, $X_d \to 0$, the NOON state has an improvement of $\sqrt{2}$ to the coherent state with the same input average photon number, or an improvement of 2 to the single photon state. In the following section VII, we propose an alternative method based on the projective measurement for the estimation of $\triangle$.

\section{VII. PROJECTIVE MEASUREMENTS TO OBTAIN ORD i.e. RELATIVE PHASE}

Though the sensitivity of phase difference $\delta \triangle$ can not be obtained from intensity
measurements for Fock states or NOON state, we are still able to study the same
through the information encoded in the density matrix $\rho(t)$. As shown in \eqref{2density} and
\eqref{3density}, only the off-diagonal terms contain the phase parameter $\triangle$. The fidelity, i.e.
the degree to which that the output state resembles the input state, defined as
$\bra{\psi_{in}}\rho_{out}\ket{\psi_{in}}$, can provide
information about $\triangle$ by virtue of the off-diagonal terms in the density matrix. For the
single-photon state, upon projecting $\rho(t)$ to the state
$\ket{1_{H},0_{V}}=\frac{1}{\sqrt{2}}(\ket{1_{+},0_{-}}+\ket{0_{+},1_{-}})$, the fidelity is calculated as 
\begin{equation}
F_{1}=\frac{1}{2}[1-X_{s}+\sqrt{(1-X_{s})^{2}-X_{d}^{2}}cos\triangle ]\,,\label{1fidelity}
\end{equation}

\noindent where the cross-terms in $\rho(t)$ contribute to a $cos\triangle t$ term in $F_{1}$, which results
in a fringe pattern. The pattern can be Fourier transformed to enable an estimation of $\Delta$. 
Similarly, we obtain the fidelity with the NOON state input,
$\ket{1_{H},1_{V}}=\frac{1}{\sqrt{2}}(\ket{2_{+},0_{-}}-\ket{0_{+},2_{-}})$, as 
\begin{equation}
F_{2}=\frac{1}{2}\{(1-X_{s})^{2}+X_{d}^{2}+[(1-X_{s})^{2}-X_{d}^{2}]cos2\triangle\}\,,\label{2fidelity}
\end{equation}

\noindent where the cross terms now contribute to a fringe pattern with double the frequency. The
sensitivity in the estimation of $\triangle$ is consequently doubled to the preceding scenario.

\section{VIII. CONCLUSIONS}

In summary, we have computed the Cram\'er-Rao bounds relevant to the estimation of chiral parameters for three different input states: a coherent state
$\ket{\alpha _{H},\beta _{V}}$, a single-photon Fock state $\ket{1_{H},0_{V}}$, and a NOON state $\ket{1_{H},1_{V}}$. Unsurprisingly, the measurement
sensitivities for
the coherent state imposed by the Cram\'er-Rao bound coincide with the precision obtained through intensity
measurements. The single-photon input state reveals a large improvement
in the measurement of the net absorption rate, compared against the coherent states. Particularly, we find that $\delta X_{s}\rightarrow 0$ in the weak-absorption regime, i.e.,
$X_{s}\rightarrow 0$. The effect is manifestly quantum. Further improvements in both the net absorption rate
$X_{s}$ and the CD absorption difference $X_{d}$ are achieved by using the NOON state. Both the
Cram\'er-Rao bounds $\delta X_{s,min}$ and $\delta X_{d,min}$ become vanishingly small for this choice of input as
$X_{s}\rightarrow 0$, implying infinite theoretical improvement in this limit. This is to be contrasted
against a coherent state with the same input photon number, for which the sensitivity $\delta X_{d}$
approaches a constant nonzero number $\sqrt{1/2}$. It is useful to note that the sensitivities from ordinary
intensity measurements also yield relatively better results for quantum sources for the estimation of $X_{s}$, which lie close to the
lowest bounds when the absorption is weak. Since such schemes are widely in practice, this gives us a
readily accessible mechanism to exploit the utilities of these light sources. However, when the absorption
rate is higher, the QFIM method would lead to a more significant improvement, allowing us to achieve better
precision by measuring the QFIM. With all this in mind, we conclude that the use of the quantum NOON state,
generated by the SPDC, would be a desirable choice to measure circular dichroism with enhanced precision.

\section{ACKNOWLEDGMENTS}

We thank the support of Air Force Office of Scientific Research (Award N%
\textsuperscript{\underline{o}} FA-9550-20-1-0366) and the Robert A Welch
Foundation (A-1943-20210327). JW thanks Debsuvra Mukhopadhyay for reading the paper and for suggestions. GSA thanks Vlad Yakovlev and Luiz Davidovich for discussions on quantum metrology and chirality.

\section{APPENDIX A: DERIVATION OF THE SLDS WITH THE COHERENT STATE INPUT}

We give a detailed derivation. For simplicity, we use a coherent state $|\beta>=e^{-\frac{1}{2}|\beta|^{2}}\sum\frac{\beta^{k}}{\sqrt{k!}}|k>$ to study a 
sample with a single absorption rate $\alpha=1-e^{-2\gamma t}$. The master equation in this case is

\begin{equation}\tag{A1}
\frac{\partial\rho}{\partial t}=-\gamma(\rho a^{+}a-2a\rho a^{+}+a^{+}a\rho)\,.\label{appendix1}
\end{equation}

\noindent Coherent states remains coherent after damping, $\rho(t)=|\beta e^{-\gamma t}><\beta e^{-\gamma t}|$. 
We write it on the Fock state bases

\begin{equation}\tag{A2}
\rho(t)=e^{-|\beta|^{2}e^{-2\gamma t}}\sum\frac{e^{-(m+n)\gamma t}\beta^{m}\beta^{*n}}{\sqrt{m!n!}}|m><n|\,,\label{appendix2}
\end{equation}

\noindent and apply $\frac{\partial\rho}{\partial\alpha}=\frac{\partial t}{\partial\alpha}\frac{\partial\rho}{\partial t}$ on the matrix element 
$\rho_{mn}$

\begin{equation}\tag{A3}
\frac{\partial\rho_{mn}}{\partial\alpha}=\frac{1}{2}\rho_{mn}[e^{2\gamma t}(n+m)-2|\beta|^{2}]\,.\label{appendix3}
\end{equation}

\noindent Comparing it with $\frac{\partial\rho}{\partial\alpha}=\frac{1}{2}(L\rho+\rho L)$, the term proportional to $(m+n)$ can be obtained by $\rho_{mp}(L_{1})_{pn}=n\rho_{mn}$ and $(L_{1})_{mp}\rho_{pn}=m\rho_{mn}$, leading to $L_{1}=e^{2\gamma t}a^{+}a$. And the other term $-|\beta|^{2}\rho_{mn}$ is a constant number acting on $\rho_{mn}$, which leads to a constant part $L_{2}=-|\beta|^{2}$ in $L$. Thus we have
\begin{equation}\tag{A4}
L=L_{1}+L_{2}=\frac{1}{1-\alpha}a^{+}a-|\beta|^{2}\,.\label{appendix4}
\end{equation}

\section{APPENDIX B. THE OFF-DIAGONAL ELEMENTS OF THE QFIM WITH THE COHERENT STATE INPUT}

The off-diagonal elements between absorption and phase shifts of the QFIM in Eq. (14) is zero. We show a derivation of of $F_{d,\triangle}$ for example. Eq. (1) reads as
\begin{equation}\tag{B1}
F_{d,\triangle}=\frac{1}{2}Tr[\rho(L_{d}L_{\triangle}+L_{\triangle}L_{d})].
\end{equation}
for $F_{d,\triangle}$, where $L_{d}$ and $L_{\triangle }$ are defined in Eq. (10) and (13). The RHS of Eq. (B1) can be calculated as
\begin{equation*}
Tr[\rho L_{d}L_{\triangle}+\rho L_{\triangle}L_{d}]=-2iTr[\rho L_{d} G_{\triangle}\rho]+2iTr[\rho L_{d} \rho G_{\triangle}]
\end{equation*}
\begin{equation*}
-2iTr[\rho G_{\triangle}\rho  L_{d}]+2iTr[\rho \rho G_{\triangle} L_{d}]
\end{equation*}
\begin{equation}\tag{B2}
=-2iTr[\rho [L_{d},G_{\triangle}]]=0,
\end{equation}
Since $Tr[ABC]=Tr[BCA]$ and $[L_{d},G_{\triangle}]=0$. This would apply also to the other three off-diagonal coefficients.


\begin{thebibliography}{99}{

\bibitem{bondurant} R. S. Bondurant, P. Kumar, J. H. Shapiro, and M. Maeda, \enquote{Degenerate four-wave mixing as a possible source of squeezed-state light,} Phys. Rev. A \textbf{30}(1), 343-353 (1984).

\bibitem{teich1989} M. C. Teich, B. E. A. Saleh, \enquote{Squeezed state of light,} Quant Optic J Eur Opt Soc B \textbf{1}(2), 153-191 (1989).

\bibitem{mandel} C. K. Hong and L. Mandel, \enquote{Theory of parametric frequency down conversion of light,} Phys. Rev. A \textbf{31}(4), 2409-2418 (1985).

\bibitem{gisin} S. Tanzilli, H. De Riedmatten, H. Tittel, H. Zbinden, P. Baldi, M. De Micheli,
D. B. Ostrowsky, and N. Gisin, \enquote{Highly efficient photon-pair source using periodically poled lithium niobate waveguide,} Electron. Lett. \textbf{37}(1), 26-28 (2001).

\bibitem{marocco} F. De Martini, G. Di Giuseppe, M. Marrocco, \enquote{Single-Mode Generation of Quantum Photon States by Excited Single Molecules in a Microcavity Trap,} Phys. Rev. Lett. \textbf{76}(6), 900-903 (1996).

%

\bibitem{michel} L. Brahim and O. Michel, \enquote{Single-photon sources,} Rep. Prog. Phys. \textbf{68}, 1129-1179 (2005).

\bibitem{beveratos} A. Beveratos, S. Kühn, R. Brouri, T. Gacoin, J.-P. Poizat and P. Grangier , \enquote{Room temperature stable single-photon source,} Eur. Phys. J. D \textbf{18}, 191-196 (2002).

\bibitem{cramer} H. Cram\'er, \enquote{Mathematical Methods of Statistics,}
(Princeton University Press, 1946, pp. 500).

\bibitem{fisher} R. A. Fisher, \enquote{On the Dominance Ratio,} Proc. R. Soc. Edinburgh \textbf{42}, 321-341
(1922).

\bibitem{helstrom} C. W. Helstrom, \enquote{Quantum Detection and Estimation
Theory,} (Academic Press, 1976, Chap. VIII.4).

%

\bibitem{holevo} A. S. Holevo, \enquote{Probabilistic and Statistical Aspects of
Quantum Theory,} (North-Holland, 1982).

\bibitem{caves2} S. L. Braunstein and C. M. Caves, \enquote{Statistical distance and the geometry of quantum states,} Phys. Rev. Lett. \textbf{%
72}, 3439-3443 (1994).

\bibitem{davi1} B. M. Escher, R. L. de Matos Filho, and L. Davidovich, \enquote{General framework for estimating the ultimate precision limit in noisy quantum-enhanced metrology,} Nat. Phys. \textbf{7}(5), 406-411 (2011).

\bibitem{davi2} C. L. Latune, B. M. Escher, R. L. de Matos Filho, and L. Davidovich, \enquote{Quantum limit for the measurement of a classical force coupled to a noisy quantum-mechanical oscillator,} Phys. Rev. A \textbf{88}, 042112 (2013).

\bibitem{davi3} B. M. Escher, L. Davidovich, N. Zagury, and R. L. de Matos Filho, \enquote{Quantum Metrological Limits via a Variational Approach,} Phys. Rev. Lett. \textbf{109}(19), 190404 (2012).

\bibitem{braun2} L. J. Fiderer, J. M. E. Fraisse, and D. Braun, \enquote{Maximal Quantum Fisher Information for Mixed States,} Phys. Rev. Lett. \textbf{123}, 250502 (2019).

\bibitem{jordan} A. N. Jordan, J. M.-Rincon and J. C. Howell, \enquote{Technical Advantages for Weak-Value Amplification: When Less Is More,} Phys. Rev. X \textbf{4}, 011031 (2014).

\bibitem{becker} W. Becker, \enquote{Advanced time-correlated single photon counting techniques,} (Springer, 2005, Chap. 2).

\bibitem{bachor} H.-A. Bachor, and T. C. Ralph, \enquote{A Guide to Experiments in Quantum Optics,} 2nd edn, (Wiley-VCH, 2004, Chap. 7).


\bibitem{greenfield} N. J. Greenfield, \enquote{Using circular dichroism spectra to estimate protein secondary structure,} Nat. protoc. \textbf{1}(6), 2876-2890
(2006).

\bibitem{whitmore} L. Whitmore, B. A. Wallace, \enquote{Protein secondary structure analyses from circular dichroism spectroscopy: methods and reference databases,} Biopolymers \textbf{89}(5), 392-400 (2008).

\bibitem{provencher} S. W. Provencher, J. Gloeckner, \enquote{Estimation of globular protein secondary structure from circular dichroism,} Biochemistry \textbf{20}(1), 33-37 (1981).

\bibitem{sreerama} N. Sreerama, R. W. Woody, \enquote{A self-consistent method for the analysis of protein secondary structure from circular dichroism,} Anal. Biochem. \textbf{209}(1), 32-44 (1993).

\bibitem{cathou68} R. E. Cathou, A. Kulczycki, E. Haber, \enquote{Structural features of gamma-immunoglobulin, antibody, and their fragments. Circular dichroism studies,} Biochemistry \textbf{7}(11), 3958-3964 (1968).

%

\bibitem{cathou70} R. E. Cathou, T. C. Werner, \enquote{Hapten stabilization of antibody conformation,} Biochemistry \textbf{9}(16), 3149-3155 (1970).

\bibitem{joshi} V. Joshi, T. Shivach, N. Yadav, and A. S. Rathore, \enquote{Circular dichroism spectroscopy as a tool for monitoring aggregation in monoclonal antibody therapeutics,} Anal. Chem.  \textbf{86}(23): 11606-11613 (2014).

\bibitem{bienkiewicz} E. A. Bienkiewicz, J. N. Adkins, K. J. Lumb, \enquote{Functional consequences of preorganized helical structure in the intrinsically disordered cell-cycle inhibitor p27(Kip1),} Biochemistry \textbf{41}(3), 752-759 (2002).

\bibitem{jennings} P. A. Jennings, and P. E. Wright, \enquote{Formation of a molten globule intermediate early in the kinetic folding pathway of apomyoglobin,} Science \textbf{262}%
(5135), 892-896 (1993).

%

\bibitem{tang} Y. Zhou, Z. Zhu, W. Huang, W. Liu, S. Wu, X. Liu, Y. Gao, W. Zhang, and Z. Tang, \enquote{Optical Coupling Between Chiral Biomolecules and Semiconductor Nanoparticles: Size-Dependent Circular Dichroism Absorption†,} Angew. Chem. Int. Ed. \textbf{50},
11456-11459 (2011).

\bibitem{bailey} J. Bailey, \enquote{Astronomical Sources of Circularly Polarized Light and the Origin of Homochirality,} Orig. Life Evol. Biosph. 
\textbf{31}, 167-183 (2001).

\bibitem{azzam} R. M. A. Azzam and N. M. Bashara,  \enquote{Ellipsometry and Polarized Light,} (North-Holland, 1977).

\bibitem{setala} A. Hannonen, A. T. Friberg, and T. Setala, \enquote{Classical spectral ghost ellipsometry,} Opt. Lett. \textbf{41}(21), 4943-4946 (2016).

\bibitem{caves} C. M. Caves, \enquote{Quantum-mechanical noise in an interferometer,} Phys. Rev. D \textbf{23}, 1693-1708 (1981).

\bibitem{kwait}  D. F. V. James, P. G. Kwiat, W. J. Munro, and A. G. White, \enquote{Measurement of qubits,} Phys. Rev. A  \textbf{64}, 052312 (2001).

\bibitem{agarwal10} U. Schilling, J. von Zanthier, and G. S. Agarwal, \enquote{Measuring arbitrary-order coherences: Tomography of single-mode multiphoton polarization-entangled states,} Phys. Rev. A  \textbf{81}, 013826 (2010).


%

\bibitem{teich1} A. F. Abouraddy, K. C. Toussaint, A. V. Sergienko, B. E. A. Saleh, and M. C. Teich, \enquote{Ellipsometric measurements by use of photon pairs generated by spontaneous parametric downconversion,} Opt. Lett. \textbf{26}(21), 1717-1719 (2001).

\bibitem{teich2} A. F. Abouraddy, K. C. Toussaint, A. V. Sergienko, B. E. A. Saleh, and M. C. Teich, \enquote{Entangled-photon ellipsometry,} JOSA B \textbf{19}(4): 656-662 (2002).

\bibitem{toussaint}  K. C. Toussaint, G. Di Giuseppe, K. J. Bycenski, A. V. Sergienko, B. E. A. Saleh, and M. C. Teich, \enquote{Quantum ellipsometry using correlated-photon beams,} Phys. Rev. A \textbf{70}(2), 023801 (2004).

\bibitem{huber} C. Czeranowsky, E. Heumann, and G. Huber, \enquote{All-solid-state continuous-wave frequency-doubled Nd:YAG–BiBO laser with 2.8-W output power at 473 nm,} Opt. Lett. \textbf{28}(6) 432-434 (2003).

\bibitem{kolkiran} Aziz Kolkiran and G. S. Agarwal, \enquote{Towards the Heisenberg limit in magnetometry with parametric down-converted photons,} Phys. Rev. A \textbf{74}, 053810 (2006).

%

\bibitem{paris} A. Monras and M. G. Paris, \enquote{Optimal Quantum Estimation of Loss in Bosonic Channels,} Phys. Rev. Lett. \textbf{98},
160401 (2007).

\bibitem{braun1} P. Binder, and D. Braun, \enquote{Quantum parameter estimation of the frequency and damping of a harmonic oscillator,} Phys. Rev. A \textbf{102}(1), 012223 (2020).

\bibitem{souza} G. Adesso, F. Dell'Anno, S. De Siena, F. Illuminati, and L. A. M. Souza, \enquote{Optimal estimation of losses at the ultimate quantum limit with non-Gaussian states,}
Phys. Rev. A \textbf{79}, 040305 (2009).

\bibitem{jiaxuan} J. Wang, L. Davidovich, and G. S. Agarwal, \enquote{Quantum sensing of open systems: Estimation of damping constants and temperature,} Phys. Rev. Research \textbf{2}(3), 033389 (2020).

%

\bibitem{chen} Y. Chen, H. Yuan, \enquote{Maximal quantum Fisher information matrix,} New J. Phys. \textbf{19}(6), 063023 (2017).

\bibitem{liu} J. Liu, H. Yuan, X. M. Lu, and X. Wang, \enquote{Quantum Fisher information matrix and multiparameter estimation,} J. Phys. A: Math. Theor. \textbf{53}(2), 023001 (2019).

\bibitem{datta} F. Albarelli, J. F. Friel, and A. Datta, \enquote{Evaluating the Holevo Cramér-Rao Bound for Multiparameter Quantum Metrology,} Phys. Rev. Lett. \textbf{123}, 200503 (2019).

\bibitem{ioannou} C. Ioannou, R. Nair, I. FernandezCorbaton, M. Gu, C. Rockstuhl, and C. Lee, \enquote{Optimal circular dichroism sensing with quantum light: Multi-parameter estimation approach,} arXiv:2008.03888, (2020).

\bibitem{crowley} P. J. D. Crowley, A. Datta, M. Barbieri, and I. A. Walmsley, \enquote{Tradeoff in simultaneous quantum-limited phase and loss estimation in interferometry,} Phys. Rev. A  \textbf{89}, 023845 (2014).
}

\end{thebibliography}
\end{document}